\documentclass[pre,superscriptaddress,twocolumn,footinbib]{revtex4-1}
\pdfoutput=1
\usepackage[colorlinks=true,urlcolor=blue]{hyperref}
\usepackage{amsfonts}
\usepackage{graphicx}
\usepackage{placeins}
\usepackage{amsmath}
\usepackage{xcolor}
\usepackage{color}
\usepackage{bm}

\definecolor{red}{rgb}{0.75,0,0}
\definecolor{blue}{rgb}{0,0,0.75}
\definecolor{green}{rgb}{0,0.5,0}

\begin{document}

\title{Activity driven orientational order in active nematic liquid crystals on an anisotropic substrate}

\author{D. J. G. Pearce}
\affiliation{Dept. of Theoretical Physics, University of Geneva, Switzerland}
\affiliation{Dept of Biochemistry, University of Geneva, Switzerland}
\begin{abstract}

We investigate the effect of an anisotropic substrate on the turbulent dynamics of a simulated two dimensional active nematic. This is introduced as an anisotropic friction and an effective anisotropic viscosity, with the orientation of the anisotropy being defined by the substrate. In this system we observe the emergence of global nematic order of topological defects that is controlled by the degree of anisotropy in the viscosity and the magnitude of the active stress. No global defect alignment is seen in passive liquid crystals with anisotropic viscosity or friction confirming that ordering is driven by the active stress. We then closely examine the active flow generated by a single defect to show that the kinetic energy of the flow is orientation dependent, resulting in a torque on the defect to align them with the anisotropy in the substrate.

\end{abstract}

\maketitle


Active nematic liquid crystals are fluids consisting of self (or mutually) propelling rod shaped particles resulting in an anisotropic fluid with broken rotational symmetry that drives itself at the microscopic scale \cite{Surrey:2001,Sanchez:2012}. This interesting combination of broken rotational symmetry and out of equilibrium, active behaviour has lead to an explosion of interest from both experimental and theoretical physics \cite{Kruse:2004,Giomi:2013,Giomi:2015,Surrey:2001,Sanchez:2012,Ramaswamy:2003}. There have been many successful experiments reproducing active nematics often utilising biological components, including microtubule kinesin suspensions \cite{Surrey:2001,Sanchez:2012,Guillamat:2017}, acto-myosin gels \cite{Schaller:2010,Schaller:2013} and elongated cells \cite{Dunkel:2013,Wioland:2016,You:2018,BlanchMercader:2018}, but also from inert components such as vibrated monolayers of granular rods \cite{Narayan:2007}. 

These systems have been shown to display a rich phenomenology depending on many factors such as the degree to which the system is driven \cite{Giomi:2015}, the confining geometry \cite{Edwards:2009}, the density \cite{You:2018} and the boundary conditions \cite{Giomi:2014}. By varying these factors it is possible to observe diverse spatiotemporal patterns including vortices \cite{Edwards:2009,Wioland:2016}, oscillating textures \cite{Schaller:2010,Keber:2014} and travelling bands \cite{Edwards:2009,Schaller:2010}. When the driving force is sufficiently high, active nematics can spontaneously nucleate many topological defects, generating flows and interacting chaotically in a regime referred to as low reynolds number, active turbulence \cite{Hemingway:2016,BlanchMercader:2018,Sanchez:2012,Giomi:2015}. These defects have been shown to exert an elastic torque on each other \cite{Vromans:2016} and experiments have indicated that long range nematic order of defects is possible in a state of active turbulence \cite{DeCamp:2015} though this hasn't been reproduced theoretically. It has been shown that the position and orientation of these defects can be influenced by the substrate on which the active nematic is placed. By changing the geometry of the substrate it is possible to reorient defects and sort them by charge \cite{Ellis:2018}, by changing the topology of the substrate it is possible to control the total number of defects and their trajectories \cite{Keber:2014}. A defect ordered active nematic has been recreated by placing a 2D active nematic on top of a passive liquid crystal that can be controlled by an external magnetic field\cite{Guillamat:2016}. When the passive liquid crystal layer is ordered into a smectic state by the magnetic field it creates a global anisotropy, defined by the orientation of the director in the passive liquid crystal. The resulting active nematic layer forms anti-parallel channels of topological defects \cite{Guillamat:2016}. When the two layers are in contact, the passive liquid crystal layer acts as an anisotropic dissipative agent for the flows generated in the driven active layer.

In this paper we explore how the introduction of anisotropic dissipative forces can lead to activity driven order in a simulated turbulent active nematic. This is done by introducing friction and viscosity coefficients that depend on a fixed substrate orientation. First we define a general viscosity and friction for a standard continuum model for active nematics in two dimensions. We then introduce the anisotropy to these quantities based on the substrate frame of reference. We observe that the active stress drives global nematic alignment of defects in the presence of anisotropic viscosity but not anisotropic friction. This global nematic order depends on the degree of anisotropy in the viscosity and the degree to which the active nematic is driven. We then support the hypothesis that this ordering is an active process by simulating passive liquid crystals with anisotropic viscosity which display no such ordering. Finally we explore the ordering mechanism by analysing the flow patterns around a single defect, noting that the energy dissipation of the active flows induce a torque on the defects.


First we must introduce the anisotropic friction and viscosity to the active nematic equations. We start from the generic form of the equations governing an incompressible active nematic which are given by:

\begin{align}
\rho\partial_tv_i &= \partial_j\sigma^{(t)}_{ij} - p\delta_{ij} - \mu_{ij}v_j \label{eq:v}\\
[\partial_t + v_i\partial_i]Q_{ij} &= \lambda S u_ij + Q_{ik}\omega_{kj} - \omega_{ik}Q_{kj} + \gamma^{-1}H_{ij} \label{eq:Q}
\end{align}

Where $\rho$ is the density, $\sigma^{(t)}_{ij}$ is the total stress tensor and the tensor $\mu_{ij}$ contains the friction coefficients. Since we are considering the incompressible limit, $\partial_iv_i = 0$ and we set $\rho=1$ everywhere. $Q_{ij} = S(n_in_j - \delta_{ij}/2)$ is the nematic tensor, $S$ is the nematic order parameter and $\lambda$ is the flow alignment parameter. The strain rate tensor is given by $u_{ij} = (\partial_iv_j + \partial_jv_i)/2$, vorticity tensor $\omega_{ij} = (\partial_iv_j-\partial_jv_i)/2$ and molecular tensor $H_{ij} = -\partial F/\partial Q_{ij}$ where $F$ is the Landau de Gennes free energy, and in two dimensions is given by:

\begin{equation}
F = \frac{K}{2}\int dA \left[ |\nabla Q|^2 + \frac{1}{\epsilon^2} trQ^2(trQ^2-1)\right] \label{eq:F}
\end{equation}

Where the parameter $\epsilon$ is a characteristic length which is proportional to the core defect radius and $K$ is the elastic constant associated with distortions in the director field.

The total stress tensor ($\sigma^{(t)}$) is the sum of elastic stresses ($\sigma^{(e)}$), viscous stresses ($\sigma^{(v)}$), and the active stress generated by the molecular motors ($\sigma^{(a)}$) controlled by parameter $\alpha$. In the general form, these stress tensors are given by:

\begin{align}
\sigma^{(e)}_{ij} &= -\lambda S H_{ij} + Q_{ik}H_{kj} - H_{ik}Q_{kj} \label{eq:sig_e}\\
\sigma^{(v)}_{ij} &= \nu_{ijkl}\partial_kv_l \label{eq:sig_v}\\
\sigma^{(a)}_{ij} &= \alpha Q_{ij} \label{eq:sig_a}
\end{align}

We introduce the anisotropy through the viscous stress tensor and the friction tensor. Since this anisotropy is defined by the substrate we must introduce an external frame of reference. Without loss of generality we assume that the high and low viscosity directions are aligned parallel to the $x$ and $y$ axes. With this condition we can assume the viscous stress tensor in two dimensions has the form:

\begin{align*}
\sigma^{(v)}_{xx} &= \nu_{0}\partial_xv_x\\
\sigma^{(v)}_{xy} &= (\nu_{1}\partial_xv_y + \nu_{2}\partial_yv_x)/2\\
\sigma^{(v)}_{yx} &= (\nu_{3}\partial_yv_x + \nu_{4}\partial_xv_y)/2\\
\sigma^{(v)}_{yy} &= \nu_{0}\partial_yv_y
\end{align*}

If at this stage we were to set all values of $\nu$ to be identical, we would obtain the normal viscous stress tensor for an isotropic fluid. We make the further simplifying assumption, allowed by symmetry for an incompressible fluid, that $\nu_y = \nu_1 = \nu_2$ and $\nu_x = \nu_3 = \nu_4$. Here we introduce the anisotropy by setting $\nu_x = \nu_0(1-\Delta\nu)$ and $\nu_y = \nu_0(1+\Delta\nu)$. This would mean that the dissipative effects of the perpendicular gradient of a flow are different depending on whether it is aligned with the $x$ or $y$ axis. It should be noted that the anisotropic viscosity that we have introduced here depends on an external frame of reference (the $x$ and $y$ axes) hence it is no longer Galilean invariant \cite{ViscosityFrictionComment}. The viscosity introduced here is that which would exist in a fully ordered incompressible nematic oriented parallel to the $x$ axis \cite{deGennes:1995}, which is similar to the substrate used by Guillamat et al. \cite{Guillamat:2016}.

The anisotropic friction tensor can be defined generally as:

\begin{equation}
\mu_{ij} = \mu^{(0)}\delta_{ij} + \mu^{(1)}_{ij}
\end{equation}

Where $\mu_0$ is the general isotropic substrate friction and $\mu_{ij}$ is the anisotropic part of the friction containing the required symmetries of the substrate. Since the substrate anisotropy is arranged to be parallel to the $x$ and $y$ axis this implies that the off diagonal components of the anisotropic friction tensor must be zero ($\mu^{(1)}_{xy}=\mu^{(1)}_{yx} = 0$), leaving just two friction coefficients. We introduce the anisotropy in a similar fashion to the viscosity and set $\mu_{xx} = \mu_0(1-\Delta\mu)$ and $\mu_{yy} = \mu_0(1+\Delta\mu)$. This formulation allows us to control the the degree of anisotropy fully by the dimensionless parameters $\Delta \nu$ and $\Delta\mu$; without loss of generality we choose $\Delta\nu \ge 0$ and $\Delta\mu \ge 0$. 

\begin{figure}[t]
\centering
\includegraphics[width=\columnwidth]{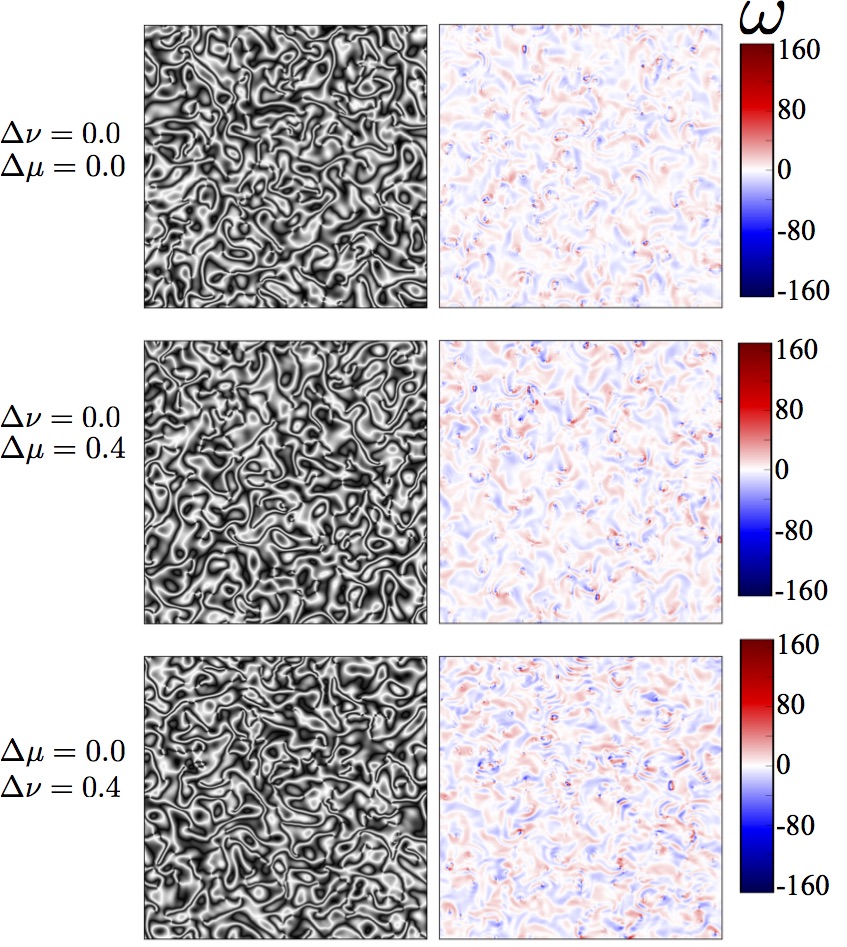}
\caption{\label{fig:active_snap} Typical director (left column) and vorticity (right column) for an isotropic active nematic (top row), an active nematic with anisotropic friction (middle row) and anisotropic viscosity (bottom row). The difference between the fields is not obvious, however some short wavelength fluctuations in the $y$ direction of the vorticity field are visible in the anisotropic viscosity case.}
\end{figure}

Equations \ref{eq:v}-\ref{eq:Q} are simulated using a stream function methodology with a periodic boundary in two dimensions. This allows us to recreate an active nematic with varying degrees of anisotropy in either the viscosity or friction. The model parameters are selected such that the system is in a state of active turbulence containing of the order of 200 defects, see S.I. for details. We choose to look at the influence of anisotropic viscosity and friction independently by either setting $\Delta\mu$ or $\Delta\nu$ to zero in all results presented here. Fig.~\ref{fig:active_snap} shows the resulting director (left column) and vorticity (right column) of active nematics with isotropic hydrodynamics (top row), anisotropic friction (middle row) and anisotropic viscosity (bottom row). From these images it is very difficult to distinguish the nematic textures of each system. The vorticity fields show some signs of anisotropy, with some short wavelength fluctuations in the $y$ direction being visible for the anisotropic viscosity system.

The lowest energy topological defects in a two dimensional nematic have half integer charge; this results in the characteristic $\pm1/2$ defects that are regularly observed in active nematics. Since these defects are not rotationally symmetric, they have an easily defined orientation which we will annotate $\psi$. The angle of the director field ($\theta$) around any of these defects can be expressed as $\theta = k(\phi - \psi) + \psi$ where $\phi$ is the polar angle between a reference axis (in this case the $x$ axis) and the position around the defect core and $k$ gives the charge of the defect \cite{Vromans:2016}. We use this definition to measure the orientation, $\psi$, of all defects in the simulated nematic. 

\begin{figure}[t]
\centering
\includegraphics[width=\columnwidth]{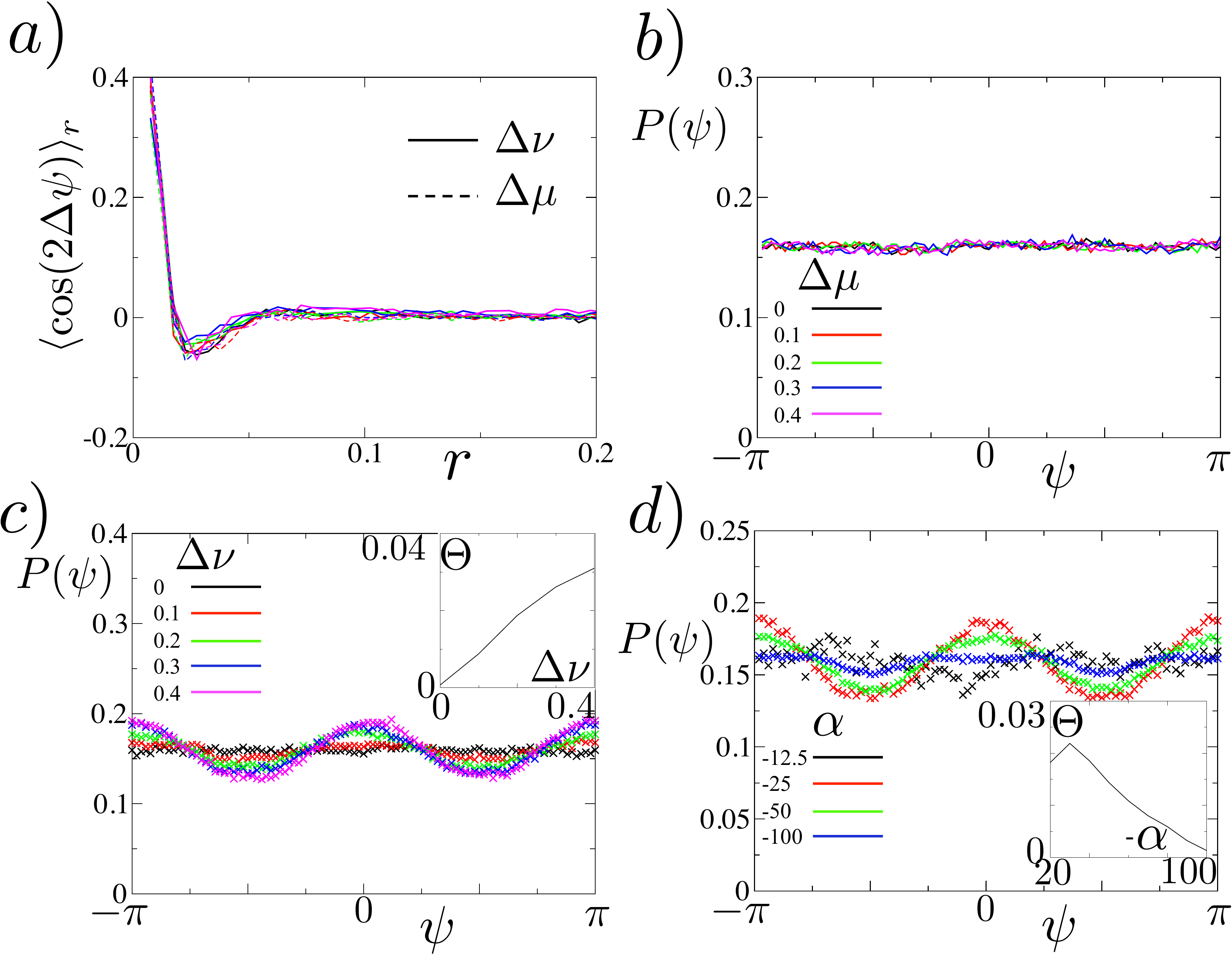}
\caption{\label{fig:active_res} (a) The nematic correlation function between positive defects. The location and depth of the minima is not significantly affected by the anisotropy, which implies that the active length scale and the elastic torques between defects are unaffected by the anisotropy. (b) The probability density function for the orientation of defects ($\psi$) within an active nematic with varying degrees of anisotropic friction ($\Delta\mu$), there is no clear order in the defect orientations. (c) The probability density function for the orientation of defects within an active nematic with varying degrees of anisotropic viscosity ($\Delta\nu$), a clear nematic order ($\Theta$) is observed that increases with the degree of anisotropy (inset). (d) The global nematic order is not observed for very small or very large values of activity for fixed anisotropy ($\Delta\nu = 0.3$). We can identify an apparent peak in the active stress (inset).}
\end{figure}

The nematic correlation function between positive defects is defined by $C_2(r) = \langle \textrm{cos}(2(\psi_i - \psi_j))\rangle_{i-j\sim r}$, shown in Fig.\ref{fig:active_res}a. Here we see that the orientational correlation length between defects is largely unaffected by the introduction of anisotropic friction or viscosity. This correlation length is set by the active length scale, $l_\alpha^2\sim K/\alpha$, which is the length at which the active and elastic forces balance and is proportional to the inter defect spacing \cite{Hemingway:2016,Giomi:2015}. The location of the minimum of the curve does not change, hence the introduction of anisotropic friction and viscosity does not affect the active length scale or the elastic torques that defects inflict upon each other at short ranges.

The distribution of $+1/2$ defect orientations within a simulated active nematic with isotropic friction and viscosity in the turbulent regime is uniform, i.e. there is no global alignment of defects. The same is true for an active nematic with anisotropic friction, Fig.~\ref{fig:active_res}b. However for an active nematic with anisotropic viscosity a clear nematic order emerges with positive defects being preferentially aligned parallel with the direction associated with the lowest viscosity, Fig.\ref{fig:active_res}c. We measure the magnitude of this order by fitting the histogram to the function $f(\psi) = 0.5/\pi + \Theta\textrm{cos}(2\psi)$ allowing us to observe that the ordering is stronger with a more significant anisotropy, see \ref{fig:active_res}c (inset). This emergent global order is mediated by the magnitude of the active stress, with the nematic ordering of the defects becoming reduced when the activity is either too high or too low, Fig.~\ref{fig:active_res}d. These results suggest that the ordering of defects within an active nematic is related to the interaction between the flow generated by the defects and the anisotropic viscosity of the fluid. However when the activity becomes very large, the alignment is lost, Fig.~\ref{fig:active_res}d(inset). This is likely due to the system becoming increasingly turbulent and chaotic, with the spacing and lifetime of defects becoming very small.


\begin{figure}[t]
\centering
\includegraphics[width=\columnwidth]{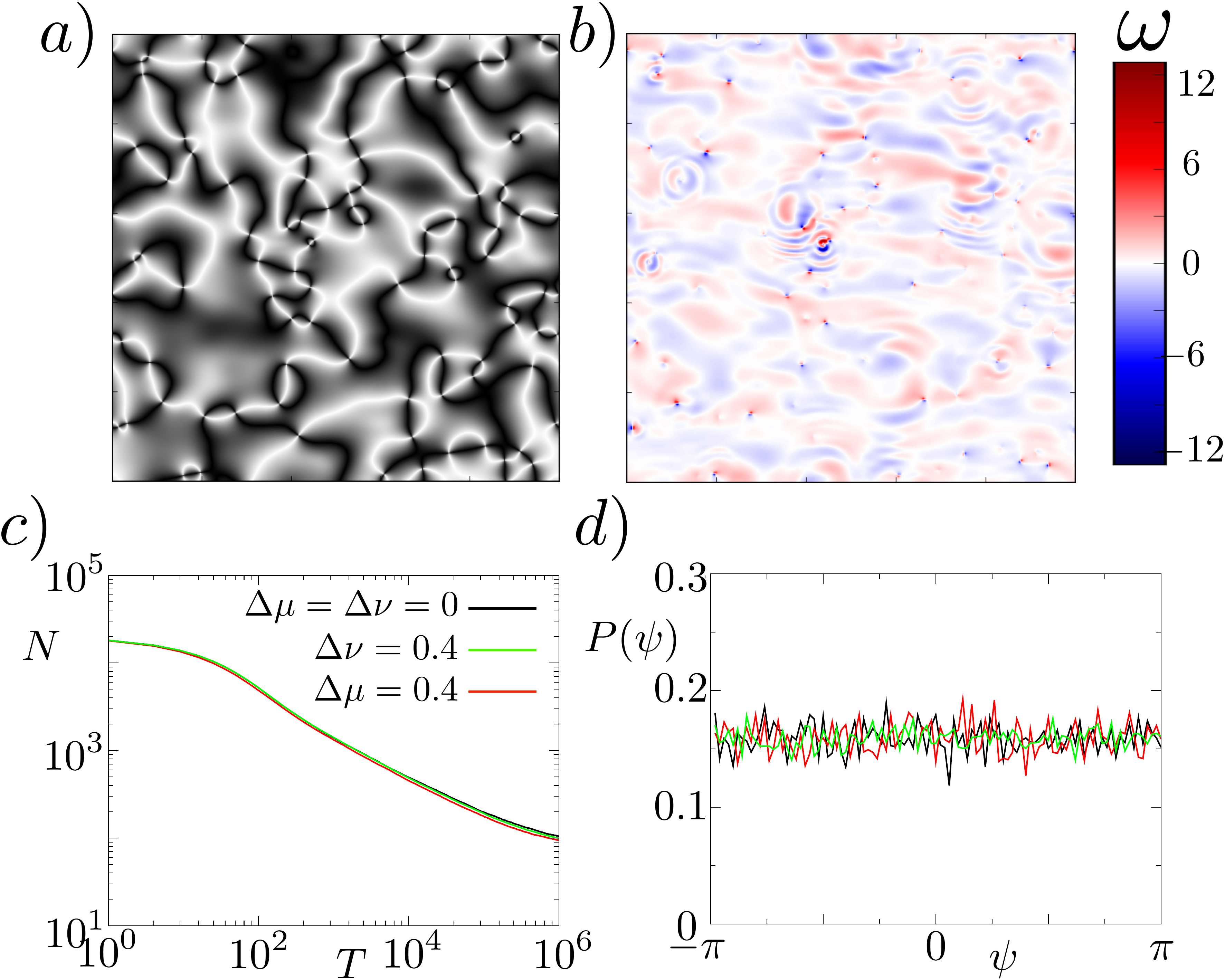}
\caption{\label{fig:passive} Snapshot of a typical nematic director (a) and vorticity (b) at the point of measurement for a passive nematic with anisotropic viscosity. (c) The number of defects as a function of simulation time. The introduction of anisotropic hydrodynamics does not appear to affect the course graining dynamics. (d) The probability density function for the orientation of positive defects in a passive nematic at the point of measurement containing 96-104 total defects. We see no emergent global order.}
\end{figure}

In order to test the hypothesis that the active flow drives a global alignment, we perform a similar study in passive nematic systems with anisotropic viscosity and friction. In the absence of an active stress, a passive nematic will relax toward the lowest energy state of the system; a uniform director field. If the nematic starts from an initially disordered state containing many defects, this process of minimising the internal energy involves significant rearrangement of the director and the annihilation of many defects, see Fig.~\ref{fig:passive}a. This motion of the nematic generates a flow in the suspending fluid, see Fig.~\ref{fig:passive}b. The introduction of anisotropic friction or viscosity does not appear to affect this relaxation process of a passive nematic, with the number of defects in all samples decaying at a very similar rate, Fig.~\ref{fig:passive}c. By simulating many passive nematics from independent, random initial conditions for the same amount of time, it is possible to create many samples of a passive nematic all at the same stage of relaxation containing a similar number of defects. This approach can be used to study large numbers of interacting defects in passive nematics, and has been used here to confirm no global nematic orientation of defects, even in cases with anisotropic friction or viscosity, see Fig.~\ref{fig:passive}d. These observations indicate that the hydrodynamic flows generated by elastic interactions alone are insufficient to generate any net defect ordering in our system.


\begin{figure}[t]
\centering
\includegraphics[width=\columnwidth]{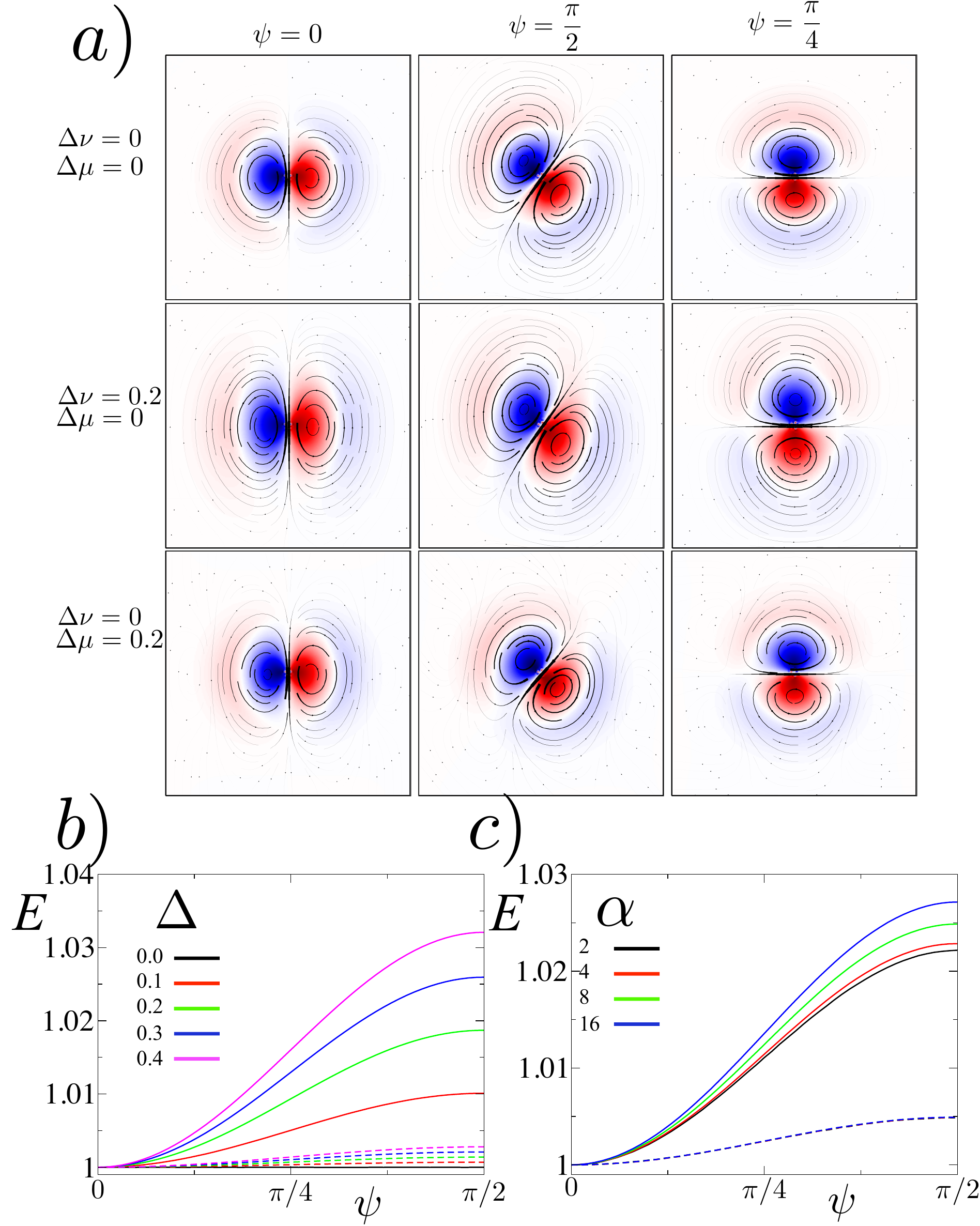}
\caption{\label{fig:single} (a) Flow pattern around a single positive defect for different orientations (column) and different anisotropies (row). We see that the mirror symmetry of flow around the defect core can be lost due to anisotropic viscosity or friction for certain orientations. (b) Kinetic energy ($E$) of the flow around each defect as a function of orientation for various degrees of anisotropy in either the viscosity (solid lines) or friction (dashed lines). (c) Kinetic energy 
of the flow around a defect for various values of activity ($\alpha$) for systems with anisotropic viscosity (solid lines) and friction (dashed lines). Lowest energy configuration is always for the defect to be parallel with the $x$ axis.}
\end{figure}

The results presented in Figs.~\ref{fig:active_res},\ref{fig:passive} indicate that the ordering of defects observed in active nematics is due to the interactions between the active flow and the anisotropic viscosity. The flow is generated by gradients in the nematic director and usually maximised around defects which generate characteristic flow patterns \cite{Giomi:2015}. The positive defects generate strong polar flows which lead them to `swim' through the fluid giving them a self propelled particle like behaviour. By simulating the flow field generated by a fixed nematic texture containing a single defect with a predetermined orientation, we can observe directly how the flow interacts with the anisotropic viscosity and friction. In an isotropic fluid, this is of course independent of the orientation of the defect, see Fig.~\ref{fig:single}a (top row). The introduction of anisotropic viscosity and friction distort these flow patterns, as they adapt to the dissipative forces of the substrate, see Fig.~\ref{fig:single}a (middle and bottom row, respectively). It is immediately apparent that when a defect is not aligned with either principal direction, the flow pattern around the defect loses its mirror symmetry in anisotropic cases, Fig.~\ref{fig:single}a (middle column). 

Fig.~\ref{fig:single}b shows the net kinetic energy of the flow around a defect $E = \rho\int v^2dA$, which for the isotropic case in independent of defect orientation. When anisotropic viscosity or friction are introduced we observe a clear dependence of the kinetic energy and the defect alignment, with the system being in a minimum energy configuration when the defect is aligned parallel to the direction of minimal shear viscosity $\psi = 0$. We see that in both cases the magnitude of the energy difference depends on the magnitude of the anisotropy $\Delta$ but is significantly larger for the anisotropic viscosity case, Fig.~\ref{fig:single}b. As the active stress is increased, we see that the dependence of the energy on the defect orientation increases for systems with anisotropic viscosity, but not in cases with anisotropic friction, \ref{fig:single}c. This supports the hypothesis that the global ordering of defects is driven by activity.


In active nematics, the active stress drives the system toward `active turbulence', a chaotic state at low reynolds number featuring many topological defects with no net order. This highlights a common feature, that the insertion of active stresses often acts to reduce order, in this case destroying the order of the nematic director and proliferating defects. When the viscosity of the active nematic has an anisotropy defined by an external frame of reference, in this case the substrate, the active flows can lead to the emergence of global nematic order of the topological defects driven by the active stress. This is evidenced by the fact that such order is not observed in systems with no active stress. Topological defects generate active flows in the fluid, the kinetic energy of which must be dissipated by the friction and viscosity of the fluid. When the fluid viscosity defined by the substrate is anisotropic, the rate of energy dissipation depends on the orientation of a defect relative to the substrate. This energy difference generates a torque on the core of the defect leading to a preferential orientation. This torque depends on the magnitude of the anisotropy and the active stress. However the active stress also defines the inter defect spacing and the defect lifetime. When the active stress is increased, the inter defect spacing is reduced. This leads to a relative increase in the elastic torques the defects exert on each other, which eventually overcomes the ordering effects of the anisotropic viscosity. In the case of anisotropic friction, the active stress does not increase the torque on the defects, so when the system is in a state of active turbulence, the disordering effects of the activity outweigh the ordering effects of the anisotropic friction in all observed cases.

\begin{acknowledgments}
I would like to thank Calres Blanch-Mercader, Karsten Kruse and Nicholas Ecker for insightful discussions.
\end{acknowledgments}


\begin{thebibliography}{99}

\bibitem{Sanchez:2012}
T. Sanchez, D. N. Chen, S. J. DeCamp, M. Heymann, Z. Dogic,
\href{http://dx.doi.org/10.1038/nature11591}{Nature {\bf 491}, 431 (2012)}.

\bibitem{Surrey:2001}
T. Surrey, F. Nedelec, S. Leibler, E. Karsenti,
\href{http://dx.doi.org/10.1126/science.1059758}{Science {\bf 292}, 1167 (2001)}.

\bibitem{Giomi:2013}
L. Giomi, M.J. Bowick, X. Ma, M.C. Marchetti,
\href{http://dx.doi.org/10.1103/PhysRevLett.110.228101}{Phys. Rev. Lett. {\bf 110}, 228101 (2013)}

\bibitem{Giomi:2015}
L. Giomi,
\href{http://dx.doi.org/10.1103/PhysRevX.5.031003}{Phys. Rev. X {\bf 5}, 031003 (2015)}

\bibitem{Kruse:2004}
K. Kruse, J.F. Joanny, F. J{\"u}licher, J. Prost, K. Sekimoto,
\href{http://dx.doi.org/10.1103/PhysRevLett.92.078101}{Phys. Rev. Lett. {\bf 92}, 078101 (2004)}

\bibitem{Ramaswamy:2003}
S. Ramaswamy, R. Aditi Simha, J. Toner,
\href{http://dx.doi.org/10.1209/epl/i2003-00346-7}{Euro Phys. Lett. {\bf 62}, 196 (2003)}.

\bibitem{Guillamat:2017},
P. Guillamat, J. Ign\'{e}s-Mullol, F. Sagu\'{e}s,
\href{http://dx.doi.org/10.1038/s41467-017-00617-1}{Nat. Commun. {\bf 8}, 564 (2017)}.

\bibitem{Schaller:2010}
V. Schaller, C. Weber, C. Semmrich,	E. Frey, A. R. Bausch,
\href{http://dx.doi.org/10.1038/nature09312}{Nature {\bf 467}, 72 (2010)}.

\bibitem{Schaller:2013}
V. Schaller, and A. R. Bausch,
\href{http://dx.doi.org/10.1073/pnas.1215368110}{Proc. Nat. Acad. Sci. U.S.A. {\bf 110}, 4488 (2013)}.

\bibitem{BlanchMercader:2018}
C. Blanch-Mercader, V. Yashunsky, S. Garcia, G. Duclos, L. Giomi, P. Silberzan,
\href{http://dx.doi.org/10.1103/PhysRevLett.120.208101}{Phys. Rev. Lett. {\bf 120}, 208101 (2018)}

\bibitem{Hemingway:2016}
E.J. Hemingway, P. Mishra, M.C. Marchetti, S.M. Fielding,
\href{http://dx.doi.org/10.1039/C6SM00812G}{Soft Matter {\bf 12}, 7943 (2016)}.

\bibitem{Dunkel:2013}
J. Dunkel, S. Heidenreich, K. Drescher, H. H. Wensink, M. B\"ar, and R. E. Goldstein,
\href{http://dx.doi.org/10.1103/PhysRevLett.110.228102}{Phys. Rev. Lett. {\bf 110}, 228102 (2013)}.

\bibitem{You:2018}
Z. You, D.J.G. Pearce, A. Sengupta, L. Giomi,
\href{https://arxiv.org/abs/1703.04504}{Phys. Rev. X {\bf}, (2018) (Accepted)}

\bibitem{Wioland:2016}
H. Wioland, E. Lushi, R. E. Goldstein,
\href{http://dx.doi.org/10.1088/1367-2630/18/7/075002}{New J. Phys. {\bf 18} 075002 (2016)}.

\bibitem{Narayan:2007}
V. Narayan, S. Ramaswamy, N. Menon,
\href{http://dx.doi.org/10.1126/science.1140414}{Science {\bf 317}, 5834 (2007)}

\bibitem{Edwards:2009}
S.A. Edwards, J.M. Yeomans,
\href{http://dx.doi.org/10.1209/0295-5075/85/18008}{Euro Phys. Lett. {\bf 85}, 18008 (2009)}.

\bibitem{Giomi:2014}
L. Giomi, A. Desimone,
\href{http://dx.doi.org/10.1103/PhysRevLett.112.147802}{Phys. Rev. Lett. {\bf 112}, 147802 (2014)}.

\bibitem{Keber:2014}
F. C. Keber, E. Loiseau, T. Sanchez, S. J. DeCamp, L. Giomi, M. J. Bowick, M. C. Marchetti, Z. Dogic, and A. R. Bausch,
\href{http://dx.doi.org/10.1126/science.1254784}{\emph{Science} {\bf 345}, 1135 (2014)}.

\bibitem{Vromans:2016}
A. Vromans, L. Giomi,
\href{http://dx.doi.org/10.1039/C6SM01146B}{Soft Matter {\bf 12}, 6490 (2016)}

\bibitem{DeCamp:2015}
S. J. DeCamp, G. S. Redner,	A. Baskaran, M. F. Hagan, Z. Dogic,
\href{http://dx.doi/org/doi:10.1038/nmat4387}{Nat. Mater. {\bf 14}, 1110 (2015)}.

\bibitem{Ellis:2018}
P. W. Ellis, D. J. G. Pearce, Y.-W. Chang, G. Goldsztein, L. Giomi, A. Fernandez-Nieves,
\href{https://dx.doi/org/10.1038/nphys4276}{Nat. Phys. {\bf 14}, 85 (2018)}.

\bibitem{Guillamat:2016}
P. Guillamat, J. Ign\'{e}s-Mullol, F. Sagu\'{e}s,
\href{http://dx.doi.org/10.1073/pnas.1600339113}{Proc. Natl. Acad. Sci. U.S.A. {\bf 113}, 5498 (2016)}.

\bibitem{ViscosityFrictionComment}
More correctly this should be referred to as a higher order friction term as it depends on the motion relative to the substrate orientation. However due to its effect on the two dimensional fluid, changing the coefficient of shear viscosity relative to orientation, we present it as an anisotropic viscosity.

\bibitem{deGennes:1995}
P. G. de Gennes, J. Prost,
{\em The Physics of Liquid Crystals},
\href{https://books.google.ch/books/about/The_Physics_of_Liquid_Crystals.html?id=0Nw-dzWz5agC&redir_esc=y}{Clarendon Press (1995)}



\end{thebibliography}
\end{document}